# MANUSCRIPT

# Significantly enhanced photocatalytic degradation of Methylene Blue using rGO-SnO$_2$ nanocomposite under natural sunlight and UV light irradiation

M. Bakhtiar Azim[a,e*], Muhammad Hasanuzzaman[b], Md. Shakhawat Hossain Firoz[c], Md. Abdul Gafur[d], A.S.W Kurny[a], Fahmida Gulshan[a*]

[a]Department of Materials and Metallurgical Engineering, Faculty of Engineering, Bangladesh University of Engineering and Technology, Dhaka-1000, Bangladesh.

[a]Professor [retired], Department of Materials and Metallurgical Engineering, Faculty of Engineering, Bangladesh University of Engineering and Technology, Dhaka-1000, Bangladesh.

[a]Professor, Department of Materials and Metallurgical Engineering, Faculty of Engineering, Bangladesh University of Engineering and Technology, Dhaka-1000, Bangladesh.

[b]Assistant Professor, Department of Glass and Ceramics Engineering, Faculty of Engineering, Bangladesh University of Engineering and Technology, Dhaka-1000, Bangladesh.

[c]Associate Professor, Department of Chemistry, Faculty of Engineering, Bangladesh University of Engineering and Technology, Dhaka-1000, Bangladesh.

[d]Pilot Plant and Process Development Center, Bangladesh Council of Scientific and Industrial Research (PP and PDC, BCSIR), Dhaka-1205, Bangladesh.

[e]School of Engineering Science, Faculty of Applied Science, Simon Fraser University, Burnaby, British Columbia, Canada.

Dr. Fahmida Gulshan, E-mail: **fahmidagulshan@mme.buet.ac.bd**

M. Bakhtiar Azim, E-mail: **bakhtiar31.mme@gmail.com; mbazim@sfu.ca**

# ABSTRACT

Environmental contamination and human exposure to dyes have dramatically increased over the past decades because of their increasing use in such industries as textiles, paper, plastics, tannery and paints. These dyes can cause deterioration in water quality by imparting color to the water and inducing the photosynthetic activity of aquatic organisms by hindering light penetration. Moreover, some of the dyes are considered carcinogenic and mutagenic for human health. Therefore, efficient treatment and removal of dyes from wastewater have attracted considerable attention in recent years. Photocatalysis, due to its mild reaction condition, high degradation, broad applied area and facile manipulation, is a promising method of solving environmental pollution problems. In this paper, we report the synthesis of reduced Graphene Oxide-Tin Oxide (rGO-$SnO_2$) nanocomposite and the effectiveness of this composite in decolorizing and degrading Methylene Blue (MB). Tin Oxide was prepared by liquid phase co-precipitation method and reduced Graphene Oxide-Tin Oxide (rGO-$SnO_2$) nanocomposite was prepared by solution mixing method. Tin Oxide ($SnO_2$) nanoparticles (NPs) have been ardently investigated as photocatalyst for water purification and environment decontamination but the photon generated electron and hole pair (EHP) recombination is one of the limiting factors. Reduced Graphene Oxide-Tin Oxide (rGO-$SnO_2$) nanocomposite is very efficient to overcome this limitation for photocatalytic application. The as-synthesized GO, $SnO_2$, GO-$SnO_2$, rGO and rGO-$SnO_2$ nanocomposite were characterized by X-ray Diffraction (XRD), Scanning Electron Microscopy (SEM), Energy Dispersive X-ray spectroscopy (EDX) and Fourier Transform Infrared spectroscopy (FTIR). The XRD data confirms the sharp peak at $2\theta=10.44°$ corresponding to (002) reflection of GO with interlayer d-spacing of 8.46 A° indication of successful preparation of GO by oxidation of graphite. Moreover, the diffraction peak shifts from 2θ=10.44° to 2θ=23.31° confirm successful synthesis of rGO as well. SEM image shows the morphology of GO, $SnO_2$, GO-$SnO_2$ and rGO-$SnO_2$. EDX studies are carried out to investigate the elemental composition and purity of the sample by giving all the elements present in the GO, $SnO_2$, GO-$SnO_2$ and rGO-$SnO_2$. The presence of functional groups was identified by FTIR. The rGO-$SnO_2$ (1:10) nanocomposite shows an efficient photodegradation efficiency of ~94% and ~95% under natural sunlight and UV light irradiation respectively for Methylene Blue (MB) within 15 minutes. Furthermore, the degradation kinetics of MB is also studied in this paper as well.

# 1. INTRODUCTION

With the rapid development of textile industry in recent years, more and more new types of dyes, such as Methylene Blue, have been produced. According to World Bank report, almost 20% of global industrial water pollution comes from the dyeing and finishing processes of textiles [1]. The discharge of azo dyes, which are stable and carcinogenic, into water bodies are harmful to human health, and cause such illness as cholera, diarrhea, hypertension, precordial pain, dizziness, fever, nausea, vomiting, abdominal pain, bladder irritation, staining of skin etc. [2]. Dyes also affect aquatic life by hindering the photosynthesis process of aquatic plants, eutrophication, and perturbation [3,4]. Numerous techniques, such as activated carbon adsorption (physical method), chlorination (chemical method), and aerobic biodegradation (biochemical method) [4] have been applied to treat textile wastewater. However, further treatments are needed, which create such secondary pollution in the environment, as the breakdown of parent cationic dyes to Benzene, $NO_2$, $CO_2$ and $SO_2$ [6]. Advanced oxidation processes (AOPs) are widely applied to mineralize dyes into $CO_2$ and $H_2O$ [7, 8]. AOPs include ozonation, photolysis, and photocatalysis with the aid of oxidants, light, and semiconductors. Photocatalytic degradation is initiated when the photocatalysts absorb photons (UV) to generate electron-hole pairs on the catalyst surface. The positive hole in the valence band ($h_{VB}^+$) will react with water to form hydroxyl radical (•OH), followed by the oxidization of pollutants to $CO_2$ and $H_2O$ [9].

Methylene Blue (MB), also known as Basic Blue 9, is a cationic azo dye (Table 1). MB is widely used in textile industries for dye processing, and upto 50% of the dyes consumed in textile industries are azo dyes [9-11]. In the past few years, several catalysts such as $TiO_2$ [6], $BiFeO_3$ [5], ZnS [13] and ZnO [9] have been used to degrade MB and the results are summarized in (Table 2).

**[Insert Table 1 and Table 2]**

Carbon based materials combined with metal oxides are known to have improved photocatalytic activity. Graphene, a two-dimensional material having $sp^2$ bonded carbon atoms arranged in a honeycomb lattice with a one-atom-thick, has recently attracted a great deal of scientific attention among the researchers. Owing to its extraordinary advantages, such as large theoretical specific surface area (2630 $m^2/g$), superior electronic and excellent chemical stability, therefore graphene is considered as an outstanding support for photocatalytic application. It was first isolated from 3D graphite by mechanical exfoliation. It has also been reported that the graphene-metal oxide composite possess good photocatalytic activity, compared to the pure metal oxide.

Numerous chemical, thermal, microwave and microbial/bacterial methods have been used in the synthesis of rGO [15]. Chemical exfoliation is preferable due to its large-scale production and low cost. Chemical exfoliation involves four steps, oxidation of graphite powder, dispersion of graphite oxide (GTO) to graphene oxide (GO), GTO exfoliation by ultrasonication to produce graphene oxide (GO) and finally, reduction of GO to rGO using a reducing agent [16]. Reduced Graphene oxide (GO) has less oxygen functional groups than graphene oxide (GO). rGO has a surface area of 833 $m^2/g$ compared to 736.6 $m^2/g$ and 400 $m^2/g$ for GO and graphite [30-31]. rGO, with its unique electronic properties, large surface area and high transparency, contributes to facile charge separation and adsorptivity in its structure. As a potential photocatalytic material, rGO-$SnO_2$ has been used in the decolorization of Methylene Blue [20] and Rhodamine B [20].

$SnO_2$ is a capable candidate to photocatalyze and complete oxidative mineralization of MB. Chemical contamination of water streams has become crucial issue of human life. Wastewater treatment plays an important role in reducing the toxic elements in wastewater. Heterogeneous photocatalysis can be applied to remove contaminants existing in wastewater effluent. Heterogeneous photocatalysis includes such reactions as organic synthesis, water splitting, photo-reduction, hydrogen transfer and metal deposition, disinfection, water treatment, removal of gaseous pollutants etc. It has become an increasingly viable technology in environmental decontamination. Photocatalytic oxidation of such organic compounds are derived by semiconductor materials like $TiO_2$, ZnO, CdS and CuO. These are now-a-days widely used in the environment as photocatalysts. They have band-gap of 3.6 eV, when other photocatalysts have higher band-gap. They are used extensively due to their low-cost, non-toxicity, high activity, large chemical stability, very low aqueous solubility and environmental friendly characteristics. Tin-assisted photocatalytic oxidation is an alternative method for purification of air and water streams. Also, in water splitting, the driving force for electrons is provided by energy of light. When $SnO_2$ is exposed to light, photocatalytic reaction is initiated. rGO-$SnO_2$ nanocomposite is very promising to overcome the limitation for photocatalytic application. rGO with its large surface area, unique electronic properties and high transparency, contributes to spatial charge separation and adsorptivity in this hybrid structure.

In this investigation, we report a facile method to prepare rGO-$SnO_2$ nanocomposite. It was synthesized via solution mixing method. The photocatalytic performances of the prepared GO, $SnO_2$, GO-$SnO_2$, rGO and rGO-$SnO_2$ nanomaterials were evaluated in the degradation of Methylene Blue (MB) under natural sunlight and UV light irradiation.

## 2. EXPERIMENTAL SECTION

### 2.1. Chemicals and Materials

Graphite fine powder (~325 mesh) was purchased from Alfa Aesar. Sodium Nitrate, Tin (II) Chloride Di-hydrate (98%), Potassium permanganate (99%), Hydrochloric acid (37%), Hydrogen Peroxide (30%), Silver Nitrate (99.8%), Ammonia solution (25%), Urea and Hydrazine Hydrate (99%) were purchased from Sigma-Aldrich (Steinheim, Germany). Sulfuric acid (98%) was obtained from Merck (Darmstadt, Germany). The chemicals were used without further purifications. Methylene Blue (MB) powder from Merck (Darmstadt, Germany) was used as the model organic dye in this study. Deionized water (DI water) was used throughout the experiments.

## 2.2. Synthesis of GO

Graphene oxide was produced through the modified Hummers' method [35] by oxidizing the graphite powder. In a typical synthesis, 5 gm of graphite powder and 2.5 gm $NaNO_3$ were mixed with 115 ml $H_2SO_4$ (conc. 98%). 15 gm of $KMnO_4$ was slowly added and stirred in an ice-bath for 1 h below 20°C. The mixture was heated to 35°C and was constantly stirred for 2 hours. The beaker was placed on an oil bath, heated to and maintained at a temperature 95°C~ 98°C, for 15 minutes. 250 ml DI water was added slowly under constant stirring. The mixture was cooled to room temperature. The beaker was then placed on the oil bath for additional 60 minutes at a constant temperature 60°C. 150 ml DI water was added under constant stirring. Finally, 50 ml (30%) $H_2O_2$ was added in drops and stirred for 2 hours. Washing, filtration and centrifugation (6000 rpm) were done until the removal of $Cl^-$ ions by using DI water. Finally, the resulting precipitate was dried at 70°C for 24 hours in an oven giving thin sheets which was Graphite Oxide (GTO). Graphite Oxide was made into fine powder form by grinding and then GTO powder was finely dispersed in DI Water. At last, ultrasonication was done for the complete exfoliation of GTO to GO.

## 2.3. Synthesis of $SnO_2$

$SnO_2$ was produced through the liquid phase co-precipitation method. 2 gm Stannous Chloride Di-hydrate ($SnCl_2.2H_2O$) was dissolved in 100 ml DI Water. After complete dissolution, ammonia solution (25%) was added in drops to the above solution under stirring. The resulting gel type precipitate was filtered and dried at 80°C for 24 hours to remove water molecules. Finally, tin oxide nanopowders were formed through calcination at 550°C for 4-6 hours.

## 2.4. Synthesis of GO-SnO$_2$

GO-SnO$_2$ was prepared through the solution mixing method. At first, 2.6 gm SnCl$_2$.2H$_2$O was added to HCl (37%) and stirred for 1 hour. Then, 260 mg GO was dispersed in 200 ml DI water by using ultrasonication for 45 minutes. Both solutions were mixed and stirred rigorously for 15 minutes. Again ultrasonication was done for 15 minutes. After that, washing, filtration and centrifugation were done with DI water till a neutral pH was obtained. The sediment was collected and dried at 80°C for 24 hours. Finally, thin sheet of GO-SnO$_2$ nanocomposite was collected and grinded into GO-SnO$_2$ nano powder.

## 2.5. Synthesis of rGO

260 mg GO was dispersed in DI water and then exfoliated by ultrasonication for 1 hour. Subsequently, 5 ml of 99% Hydrazine Hydrate was added and the solution was heated in heating mantle with stirrer at 100°C under a water cooled condenser for 24 hours. After heating at 100°C, the solution turns from brown to black precipitate. The product was cooled, filtered and washed with DI water for 4 times. Finally, the product was dried overnight at 60°C.

## 2.6. Synthesis of rGO-SnO$_2$

rGO-SnO$_2$ was prepared also through the solution mixing method. At first, 2.6 gm SnCl$_2$.2H$_2$O was added to HCl (37%) and diluted with DI water to give SnCl$_2$-HCl solution. Then, 260 mg GO was dispersed in DI water by using ultrasonication for 30 minutes to form a colloidal suspension. The GO solution was added to a SnCl$_2$-HCl solution and ultrasonication was done for 15 minutes. 2.6 gm Urea was then added to the resulting solution and stirred for 15 minutes. The solution was heated to 95°C for 6 hours. The precipitate obtained was separated by centrifugation (6000 rpm) and subsequent washing to remove excess Cl⁻ ions. The resulting product was dried in an oven at 110°C for 6 hours.

## 3. CHARACTERIZATION

The X-ray diffraction pattern of GO, $SnO_2$, GO-$SnO_2$, rGO and rGO-$SnO_2$ was recorded by a Bruker, D8 Advance diffractometer (Germany). The sample was scanned from 5° to 80° using Cu K$\alpha$ radiation source ($\lambda$ = 1.5406 A°) at 40 kV and 30 mA with a scanning speed of 0.01°s⁻1. The surface morphology of GO, $SnO_2$, GO-$SnO_2$ and rGO-$SnO_2$ was observed by FESEM-JEOL (FEG-XL 30S) Field Emission Scanning Electron microscope (FESEM). FTIR spectra of GO, $SnO_2$, and rGO-$SnO_2$ was recorded by Agilent Cary 670 FTIR spectrometer. The photodegradation percentage of MB was determined by using an ultraviolet-visible spectrophotometer (Shimadzu-UV-1601) at $\lambda_{max}$ = 664 nm and wavelength region between 400 and 800 nm. DI water was used as a reference material.

## 4. PHOTOCATALYTIC REACTION

Photocatalytic experiments were carried out by photodegrading MB using Shimadzu-UV-1601 UV-Vis spectroscopy. The solution of MB (pH~7) without rGO-$SnO_2$ was left in a dark place for 24 hrs. Then, the dye solution was exposed to natural sunlight irradiation and UV irradiation and there was no decrease in the concentration of dye. In a typical experiment, 7.5 mg of rGO-$SnO_2$ was added into a 50 mL 0.05 mM MB solution. Before illumination, the suspensions were continuously shacked and kept in dark for 60 min to reach an adsorption-desorption equilibrium between the photocatalyst and dye solution. The suspensions were then exposed to natural sunlight and UV irradiation. Samples were taken from irradiation at regular time intervals (0 min, 10 min, 15 min, 30 min, 45 min and 60 min for MB) filtered out to remove the rGO-SnO$_2$. Irradiation was carried out in falcon tubes. The same procedure was repeated for both GO, SnO$_2$, rGO and GO-SnO$_2$. The initial pH of solution (pH~7) was adjusted by small amount of 0.1 M NaOH and 0.1 M HCl when required.

Photodegradation was also observed for 0.05 mM 50 ml MB solution using only GO, SnO$_2$, GO-SnO$_2$ and rGO. The photodegradation efficiency of MB was determined by using the equation shown below:

**Photodegradation efficiency (%) = [($C_0$ - $C_t$) / $C_0$] × 100% = [($A_0$ - $A_t$) / $A_0$] × 100%** ---------------(1)

[According to 'Beer-Lambert Law']

Where, $C_0$ is the initial concentration of MB, $C_t$ is the concentration of MB at time, t and $A_0$ is the initial absorbance of MB, $A_t$ is the absorbance of at time, t.

# 5. RESULTS AND DISCUSSION

## 5.1. Characterization of Photocatalysts

### XRD

The powder X-ray diffraction pattern of GO shows a broadened diffraction peak (Fig. 1) at around $2\theta=10.44°$, which corresponds to the (002) reflection of stacked GO sheets with interlayer d-spacing of 8.46°. rGO shows diffraction peak at around $2\theta=23.31°$ with interlayer d-spacing of 7.62 A°. XRD patterns of SnO$_2$ nanoparticles (Fig. 1) shows the diffraction peaks of (110), (101),

(111) and (211) at 2θ of 26.8°, 33.9°, 37.9° and 51.8° respectively which matches well with JCPDS card # 41-1445. The average crystallite size can be determined using "Scherrer Formula":

$$B = (k\lambda)/(\beta cos\theta_\beta) \quad \text{----------------------------------------------------(2)}$$

Where, B is the average crystallite size (nm), k is the factor of 0.9, λ is the wavelength of radiation source used is 1.5406 A°, β is the full width of half maximum peak (FWHM), θ$_\beta$ is the angle at maximum peak. The average crystallite size of SnO$_2$ is ~35 nm (eq. 2). The XRD pattern of rGO-SnO$_2$ nanocomposite (Fig. 1) shows SnO$_2$ peaks but the intensity of the peaks are reduced and broaden when compared with XRD pattern of individual rGO and SnO$_2$.

**[Insert Fig. 1]**

**SEM and EDX**

SEM images of rGO-SnO$_2$ structure are shown in ((Fig. 2(e)). The average particle size of SnO$_2$ is ~46 nm using ImageJ analysis ((Fig. 2(b)). SEM images of rGO-SnO$_2$ shows presence of SnO$_2$ nanoparticles in the rGO layer by layer sheets.

**[Insert Fig. 2]**

EDX results indicating successful incorporation of oxygen by modified Hummers method in graphite layers and formation of oxygen based functional groups in GO with ratio of C/O=4.58 ((Fig. 3(a)). It also shows successful synthesis of SnO$_2$ nanoparticles ((Fig. 3(b)). It gives clear indication of incorporation of SnO$_2$ nanoparticles in rGO layer by layer sheets ((Fig. 3(d)).

**[Insert Fig. 3]**

## FTIR

The FTIR spectra of GO, $SnO_2$ and rGO-$SnO_2$ was recorded by Agilent Cary 670 FTIR spectrometer. FTIR analysis of GO shows broad absorption spectrum observed at ~3420 $cm^{-1}$ corresponding O-H stretching vibration indicating existence of absorbed water molecules and structural O-H groups in GO. The broad peak appeared in GO spectrum depicted the presence of O-H and C-H stretching. Besides, a band at 1747 $cm^{-1}$ might be related to not only the C=O stretching motion of -COOH groups situated at the edges and defects of GO lamellae but also that of ketone or quinone groups. The peak near 1700-1550 $cm^{-1}$ widens and moves to 1565 $cm^{-1}$ that reflects the presence of un-oxidized aromatic regions. The FTIR spectrum of Tin Oxide shows broad not sharp absorption band ~3420 $cm^{-1}$ corresponding O-H stretching vibration due to absorbed water molecules. A band at ~612 $cm^{-1}$ related to O-Sn-O stretching. Broad and sharp absorption spectrum observed at ~3420 $cm^{-1}$ corresponding O-H stretching vibration in rGO-$SnO_2$. Peak at 1625$cm^{-1}$ corresponding O-H bending vibration indicating existence of absorbed water molecules and structural O-H groups in rGO-$SnO_2$ by FTIR. Similarly, a sharp band at ~612 $cm^{-1}$ related to O-Sn-O stretching in rGO-$SnO_2$ refers to the incorporation of $SnO_2$ particles. (Fig. 4)

**[Insert Fig. 4]**

### 5.2. Photocatalytic Activity

5.2.1. Conditions and Parameters for Photocatalytic Activity Measurement

The concentration and amount of the dye used for the experiment were 0.05 mM and 50 ml respectively. Weight of the samples used for the experiment was 7.5 mg each. For the experiment pH was considered as ~7 and temperature outside was 29 °C - 33° C. The most crucial parameters of the experiment were (i) Natural Sunlight Irradiation, (ii) UV light Irradiation, (iii) Irradiation Time Effect and (iv) Catalyst Types.

### 5.2.2. Photocatalytic Activity Evaluation

UV-Vis Spectroscopy was used to measure absorbance of the dye solution at regular time intervals using rGO-$SnO_2$. Controlled experiments were also carried out to confirm that the degradation of MB by UV-Vis for visible range. Experiments were repeated for other photocatalysts also.

**Photodegradation Efficiency (%) = [( $C_0$ - $C_t$) / $C_0$] × 100%= [( $A_0$ - $A_t$) / $A_0$] × 100%-------------(3)**

Here, $C_0$= Concentration of MB at '0' min, $C_t$= Concentration of MB at 't' min, $A_0$= Absorbance of MB at '0' min, $A_t$= Absorbance of MB at 't' min

Time-dependent absorption spectra of MB solution shows enhanced photocatalytic activity of rGO-$SnO_2$ compared to $SnO_2$ and rGO under natural sunlight and UV irradiation (Fig. 5 and Fig 7).

### [Insert Fig. 5 and Fig. 6]

### [Insert Fig. 7 and Fig. 8]

It is clearly seen from the ((Fig. 6(a) and Fig. 8(a)) that rGO-$SnO_2$ having lesser absorbance compared to GO, $SnO_2$, GO-$SnO_2$ and rGO. From ((Fig. 6(b) and Fig. 8(b)) with irradiation time concentration of MB solution reduces significantly by using rGO-$SnO_2$ nanocomposite with respect to others under natural sunlight and UV light irradiation. Under natural sunlight and UV light irradiation rGO-SnO$_2$ nanocomposite showed ~94% and ~95% photodegradation efficiency respectively within 15 minutes compared to other photocatalysts ((Fig. 9(a) and Fig. 9(b)). The most outstanding factor of this experiment is under UV light irradiation the degradation efficiency of GO-SnO$_2$ has increased remarkably from ~61% to 94% after 30 minutes ((Fig. 9(a) and Fig. 9(b)).

**[Insert Fig. 9]**

## 6. ADSORPTION KINETICS

### 6.1. Pseudo First Order

The MB photo degradation was fitted to pseudo-first order kinetics by referring to the Langmuir-Hinshelwood kinetic model :

$$\ln\left(\frac{C_0}{C_t}\right) = k_1 t \quad \text{(4)}$$

Where, $C_t$ = Concentration of MB at time, t, $C_0$ = Initial concentration of MB, $k_1$ = Pseudo-first order rate constant.

The $k_1$ values of respective concentrations was determined from knowing the values $C_t$ and was listed in Table 3.

### 6.2. Pseudo Second Order

Pseudo second-order equilibrium adsorption equation can be expressed as follows:

$$\frac{1}{C_t} = \frac{1}{C_0} + 2k_2 t \quad \text{(5)}$$

Where, $C_t$ = Concentration of MB at time t, $C_0$ = Initial concentration of MB, $k_2$ = Pseudo-second order rate constant.

The $k_2$ values of respective concentrations was determined from knowing the values $C_t$ and was listed in Table 4.

### 6.3. Adsorption Kinetics Evaluation

The correlation coefficient ($R^2$) values are close to 1 observed in Table 3 than Table 4. Therefore, it gives clear indication that rGO-SnO$_2$ obeys the pseudo-first order kinetic model under natural sunlight irradiation.

**[Insert Table 3 and Table 4]**

**[Insert Fig. 10]**

The correlation coefficient ($R^2$) values are close to 1 observed in Table 5 than Table 6. Therefore, it gives clear indication that rGO-SnO$_2$ obeys the pseudo-first order kinetic model under UV light irradiation.

**[Insert Table 5 and Table 6]**

**[Insert Fig. 11]**

### 7. CONCLUSIONS

Results of characterization showed that GO, SnO$_2$, GO-SnO$_2$, rGO and rGO-SnO$_2$ were synthesized successfully. Degradation of 0.05 mM 50 ml Methylene Blue under direct sunlight and UV light irradiation with 7.5 mg rGO-SnO$_2$ nanocomposite as a photocatalyst takes around 15 min for ~94% and ~95% degradation respectively compared to other catalysts. The most significant result of this experiment is under UV light degradation efficiency of GO-SnO$_2$ enhanced remarkably compared to other photocatalysts within 30 minutes (from ~61% to ~94%). Wide band gap of SnO$_2$ is tuned using rGO therefore it shows enhanced photocatalytic performance compared to SnO$_2$ only. Moreover, the degradation kinetics is also studied and rGO-SnO$_2$ maintains the pseudo-first order kinetic model.

## CONFLICTS OF INTERESTS

There are no conflicts to declare.

## ACKNOWLEDGEMENTS

This work was supported by Department of Materials and Metallurgical Engineering (MME, BUET), The Pilot Plant and Process Development Center, Bangladesh Council of Scientific and Industrial Research (PP and PDC, BCSIR), Department of Glass and Ceramics Engineering (GCE, BUET) and Department of Chemistry, BUET.

# FIGURES

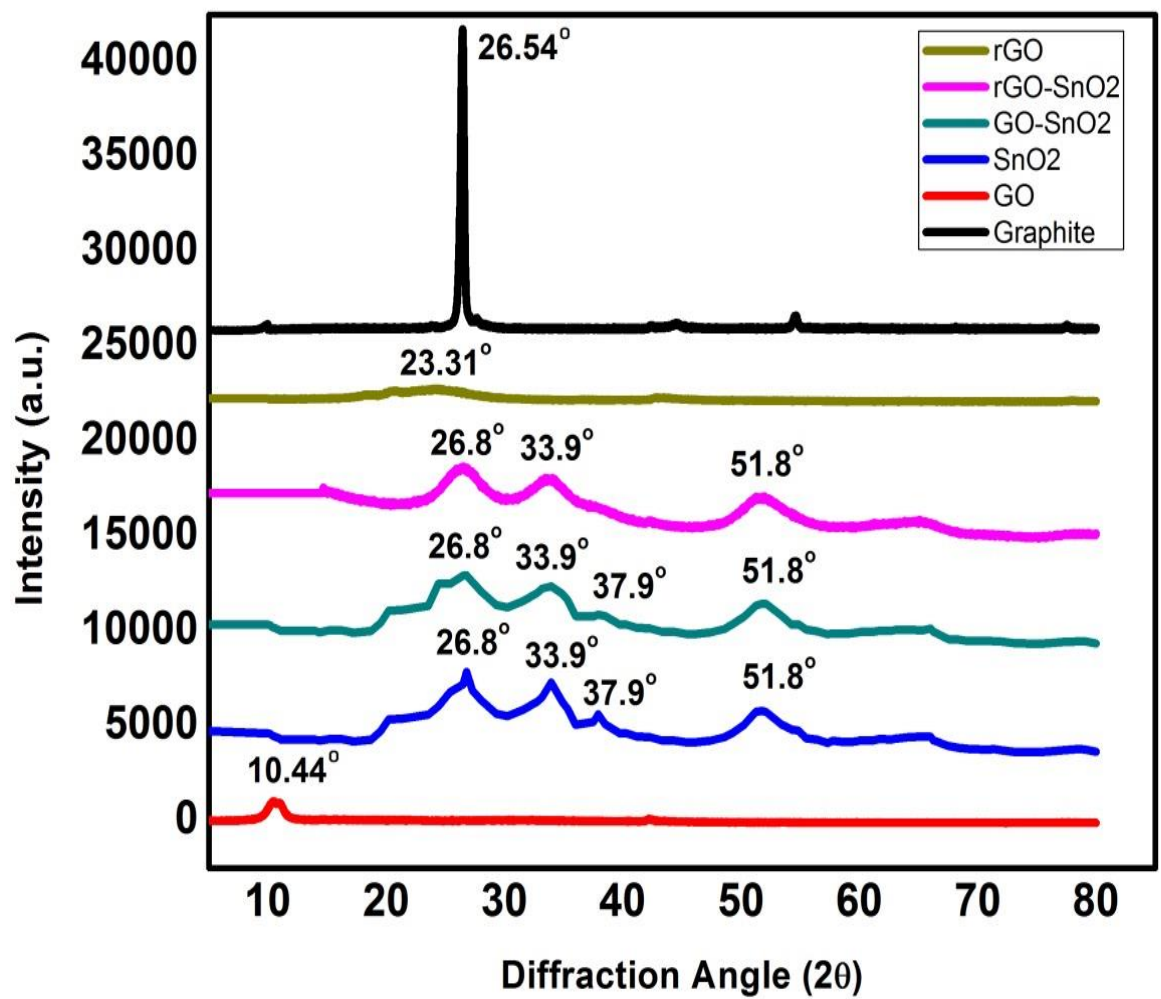

Fig. 1: XRD of Graphite, GO, SnO$_2$, GO-SnO$_2$ and rGO-SnO$_2$

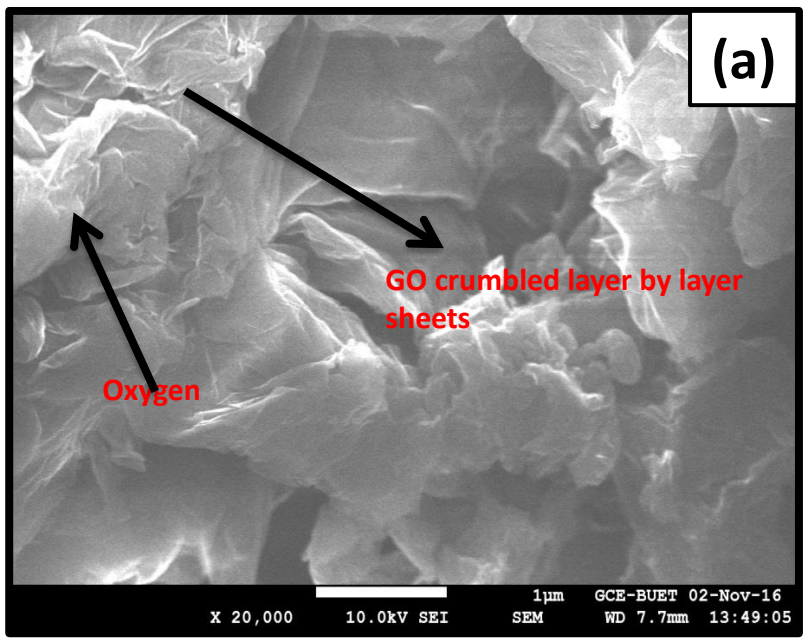

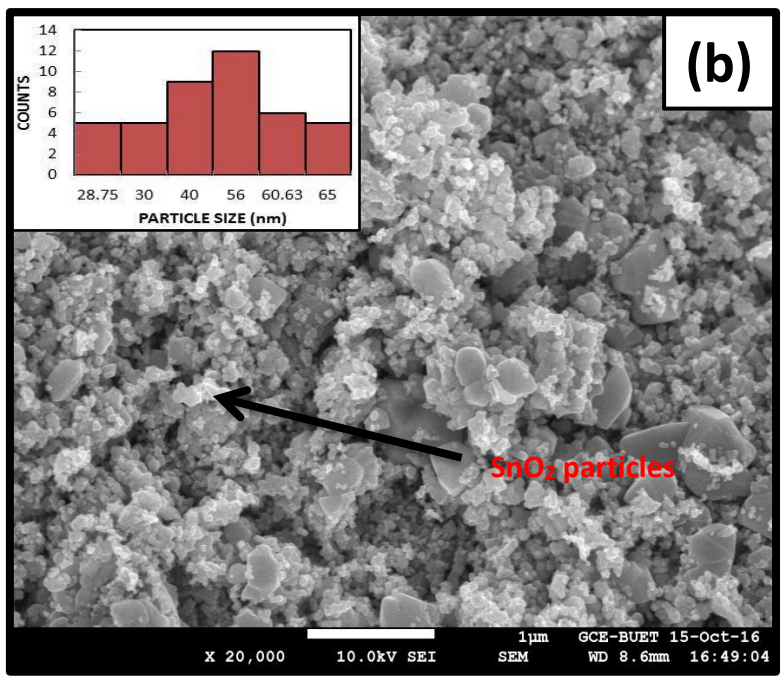

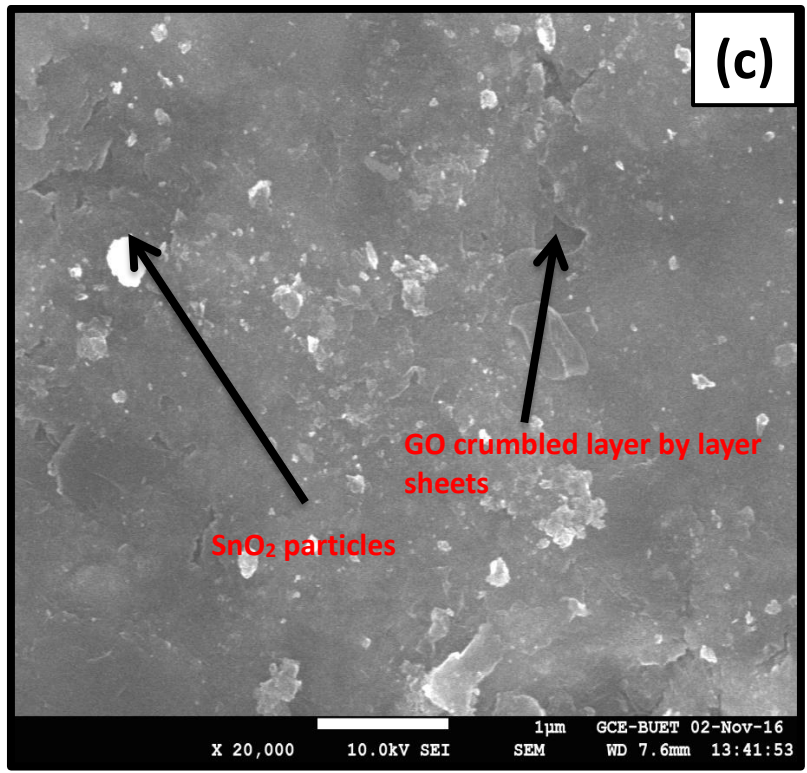

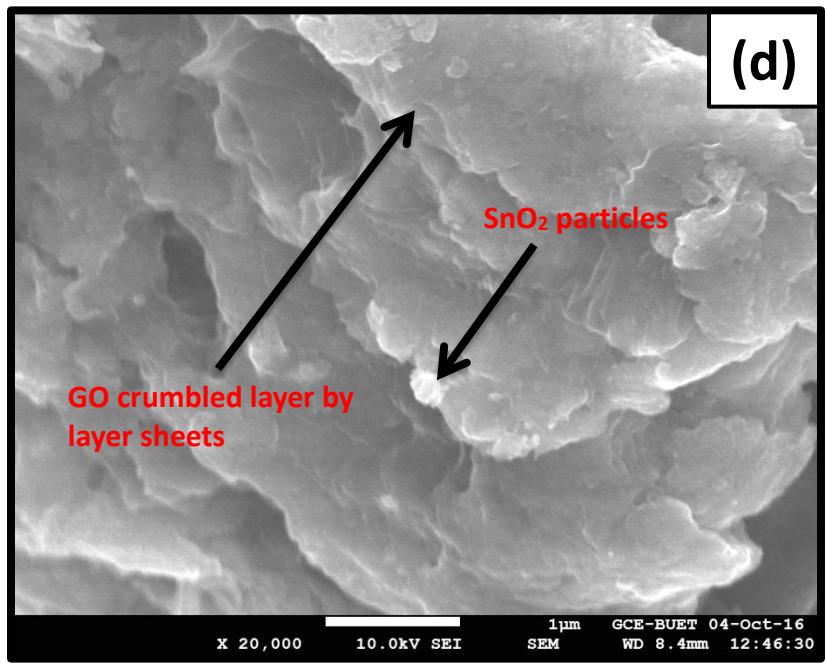

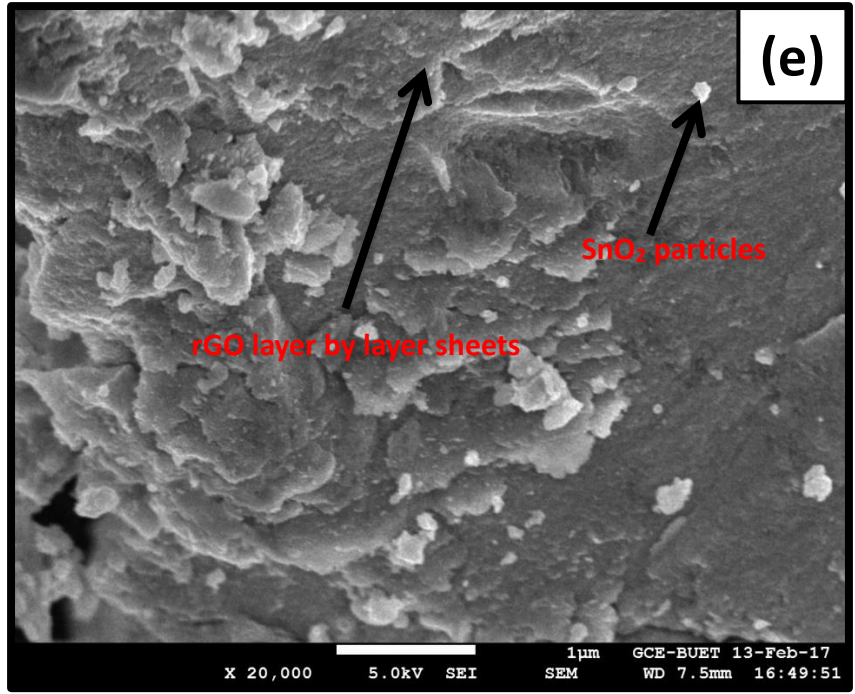

Fig. 2: SEM of (a) GO, (b) SnO$_2$, (c,d) GO-SnO$_2$ and (e) rGO-SnO$_2$

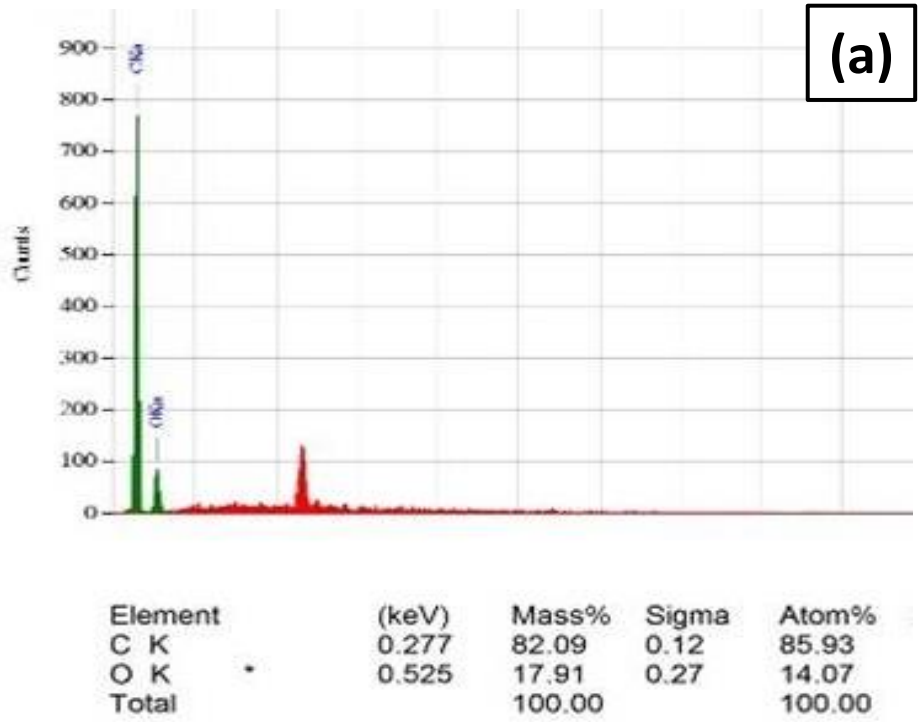

(a)

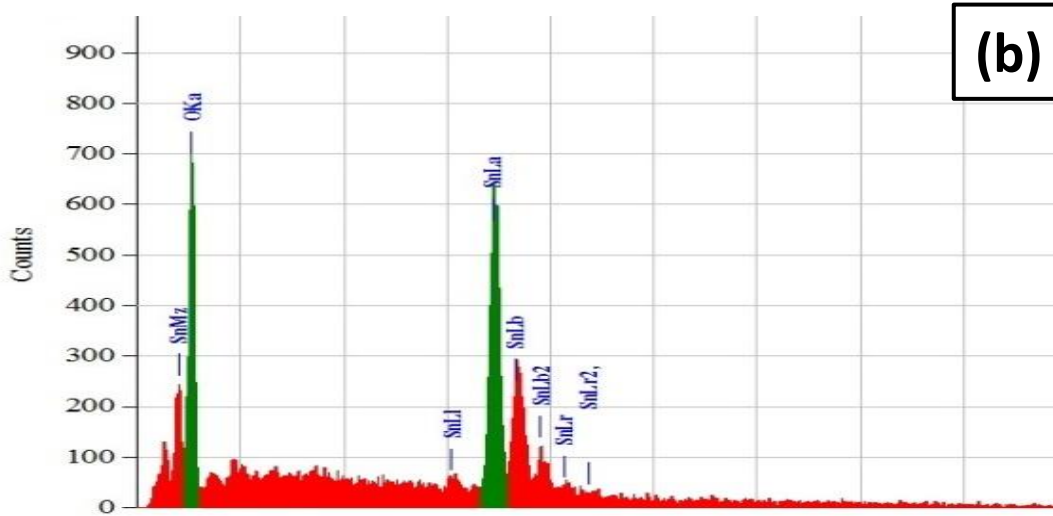

| Element | (keV) | Mass% | Sigma | Atom% |
|---|---|---|---|---|
| O K | 0.525 | 30.21 | 0.43 | 76.25 |
| Sn L | 3.442 | 69.79 | 0.94 | 23.75 |
| Total | | 100.00 | | 100.00 |

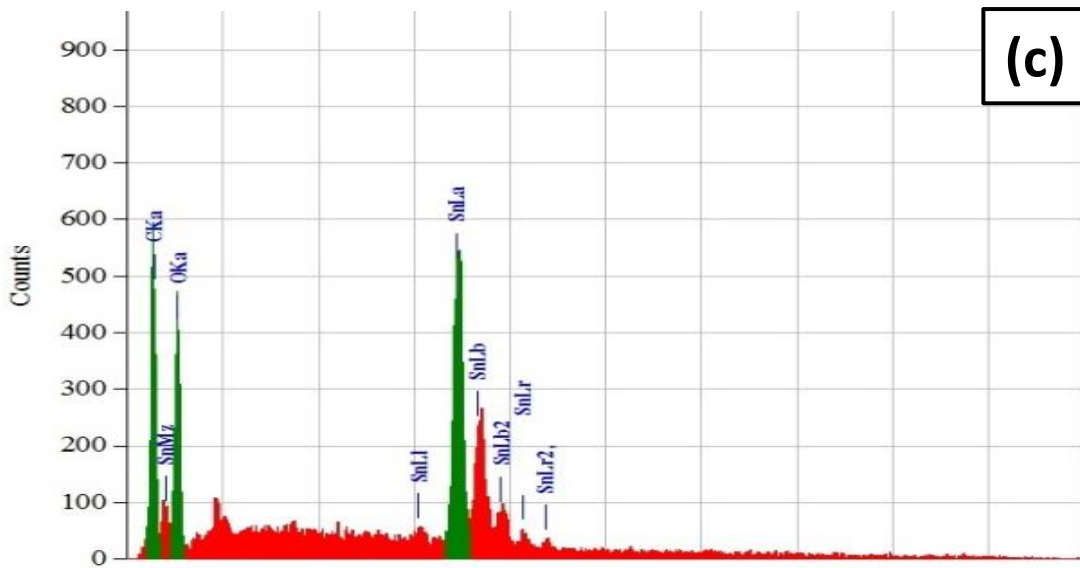

| Element | | (keV) | Mass% | Sigma | Atom% |
|---|---|---|---|---|---|
| C K | * | 0.277 | 15.05 | 0.03 | 41.42 |
| O K | | 0.525 | 19.54 | 0.14 | 40.37 |
| Sn L | | 3.442 | 65.40 | 0.54 | 18.21 |
| Total | | | 100.00 | | 100.00 |

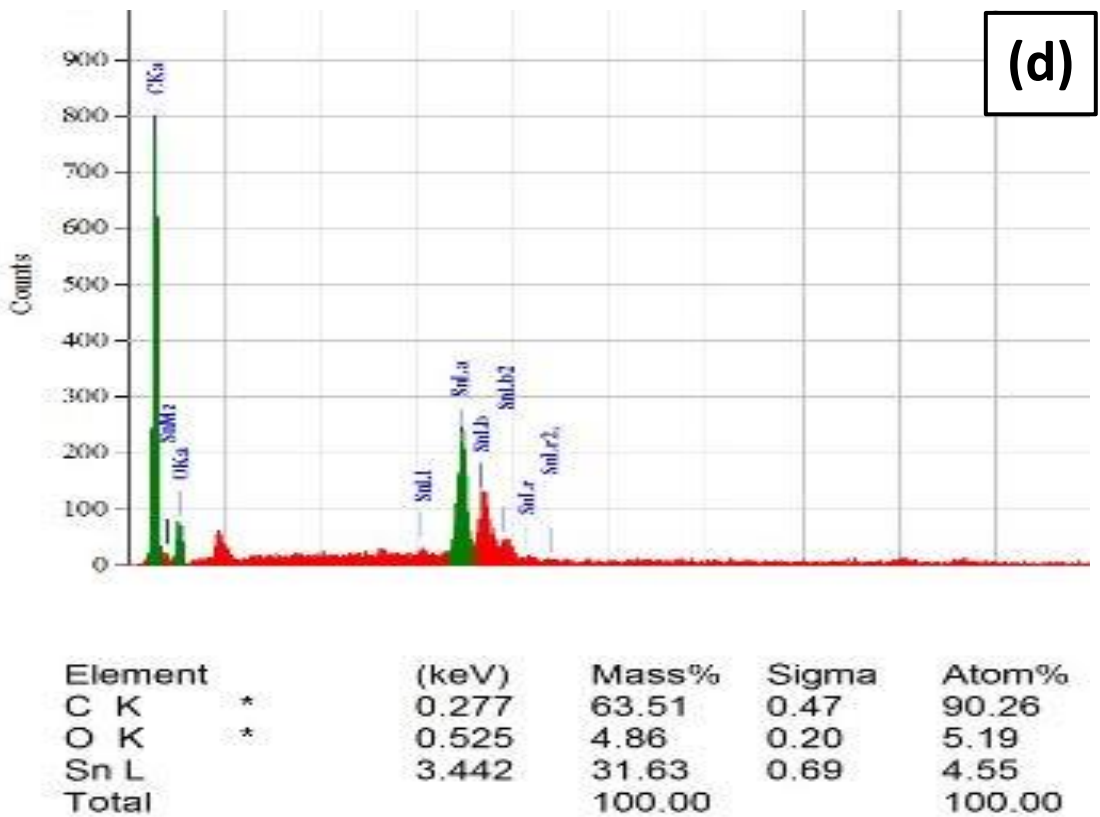

Fig. 3: EDX spectra of (a) GO, (b) SnO$_2$, (c) GO-SnO$_2$ and (d) rGO-SnO$_2$

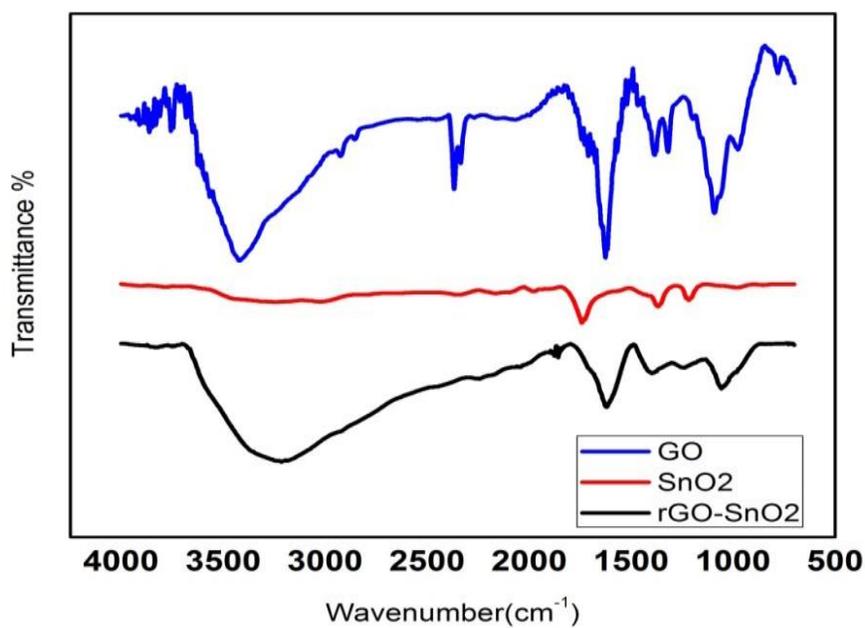

Fig. 4: FTIR curves of GO, SnO$_2$ and rGO-SnO$_2$

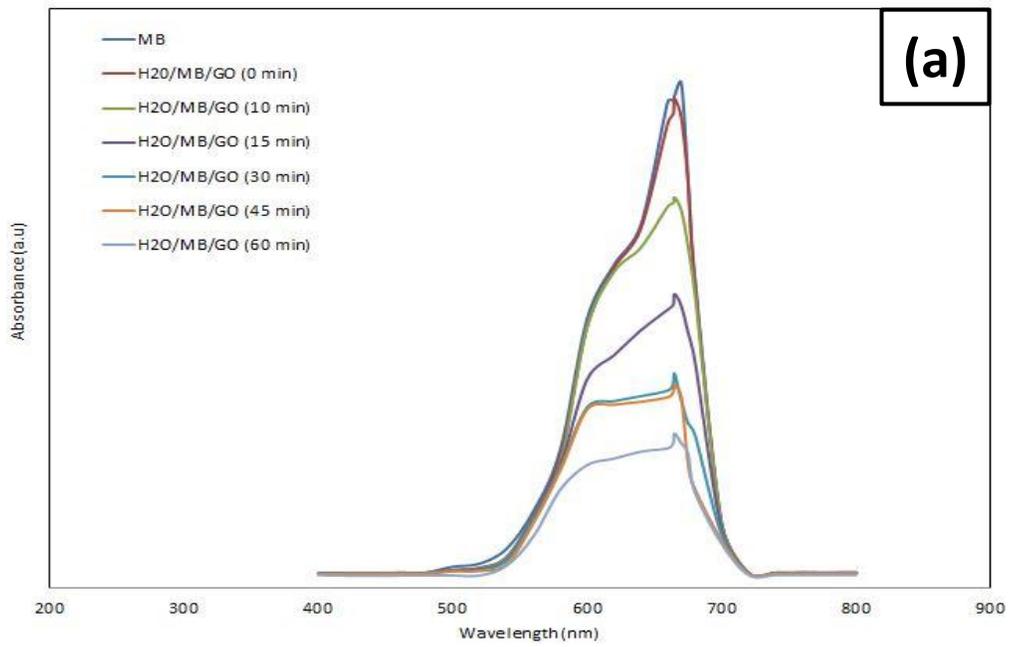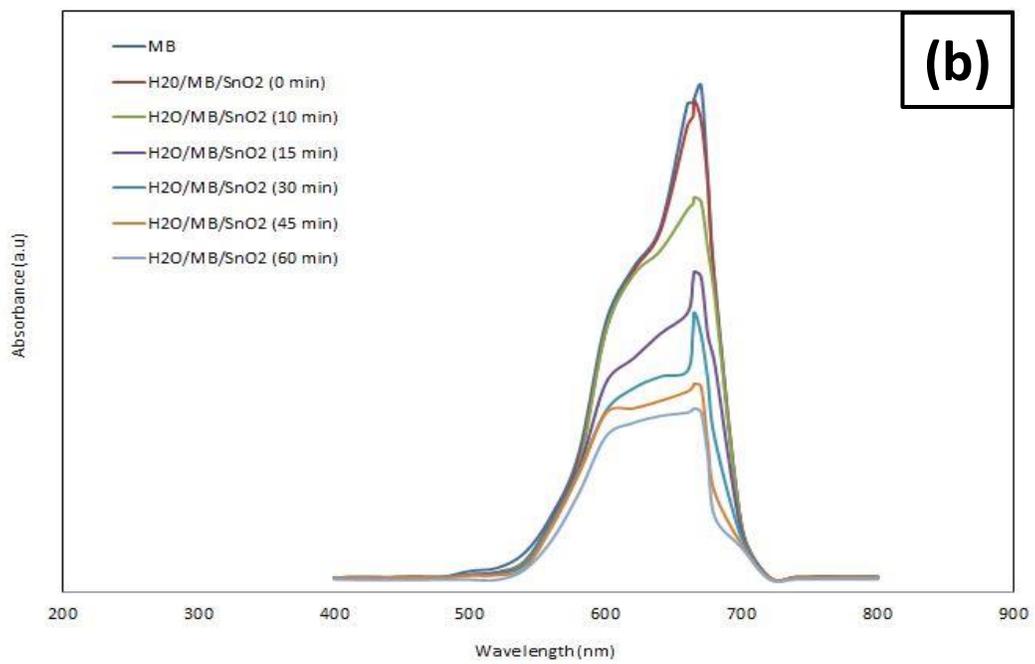

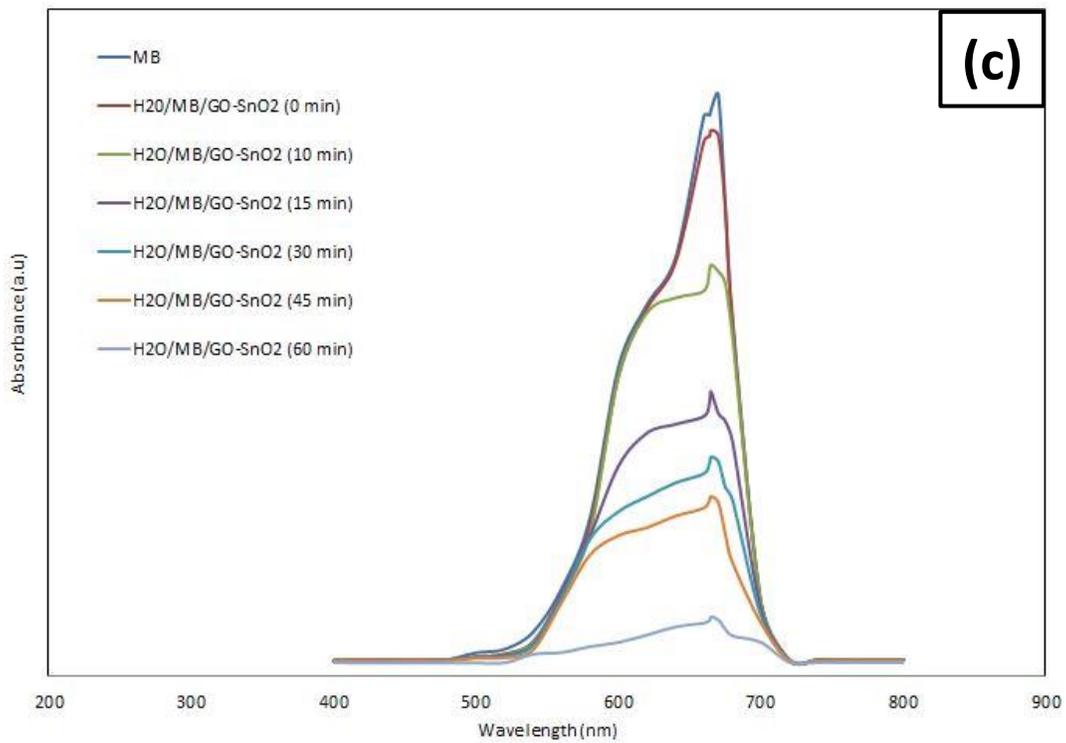

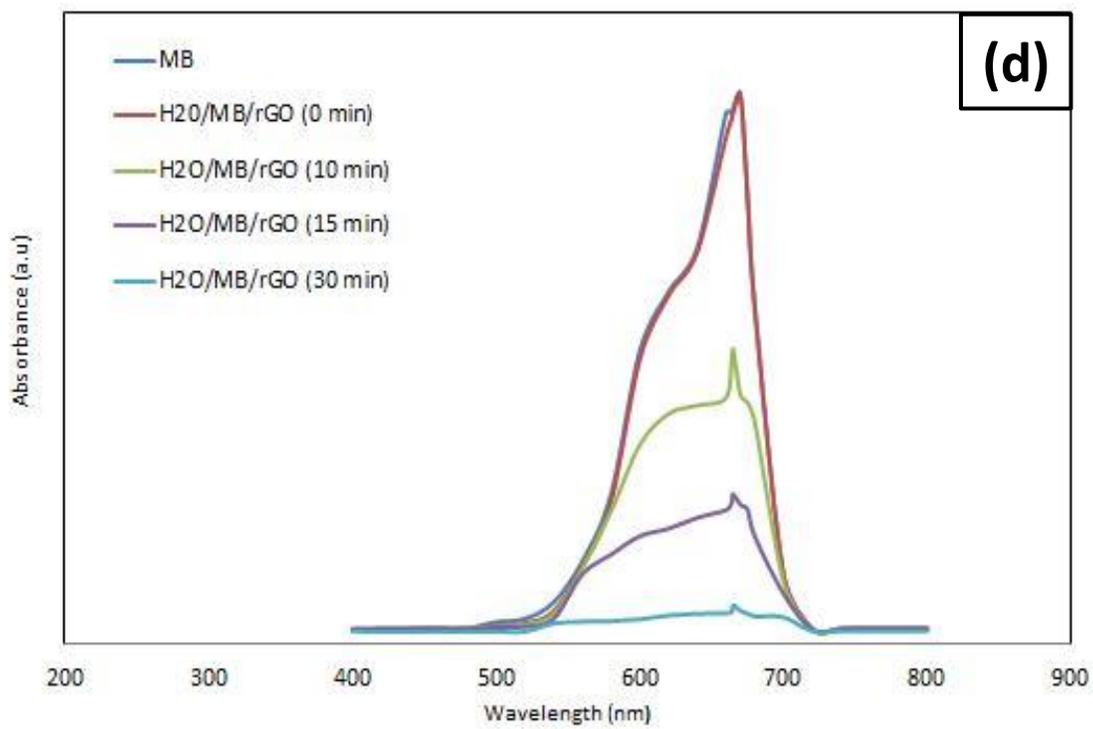

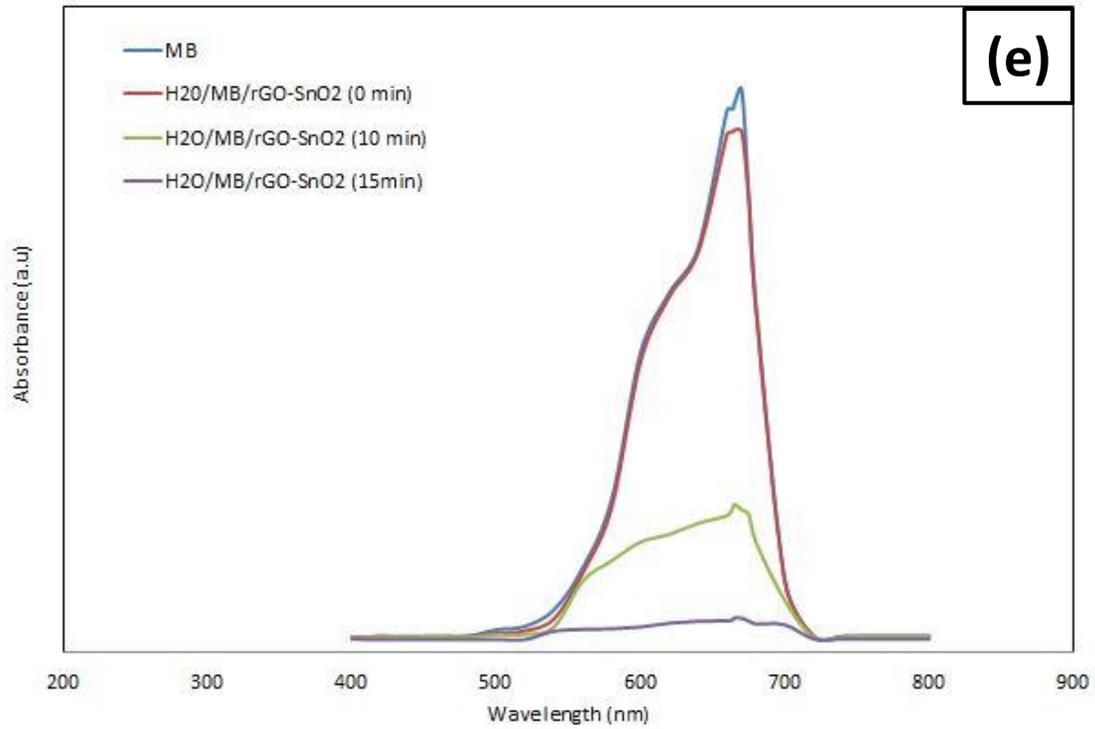

Fig. 5: Time-dependent absorption spectra of MB solution during natural sunlight irradiation in the presence (a) GO, (b) $SnO_2$, (c) GO-$SnO_2$, (d) rGO and (e) rGO-$SnO_2$

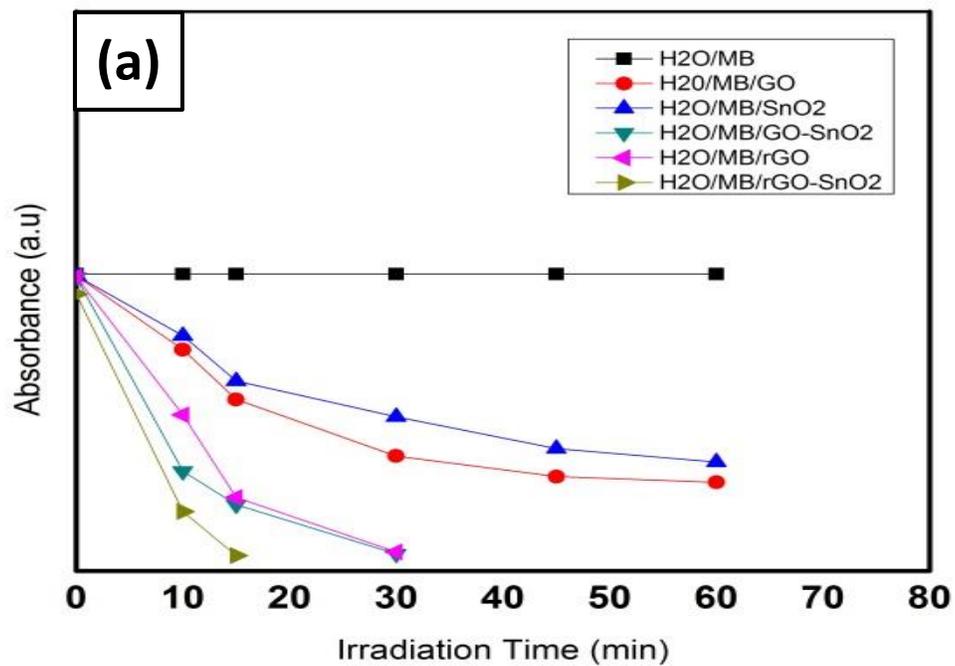

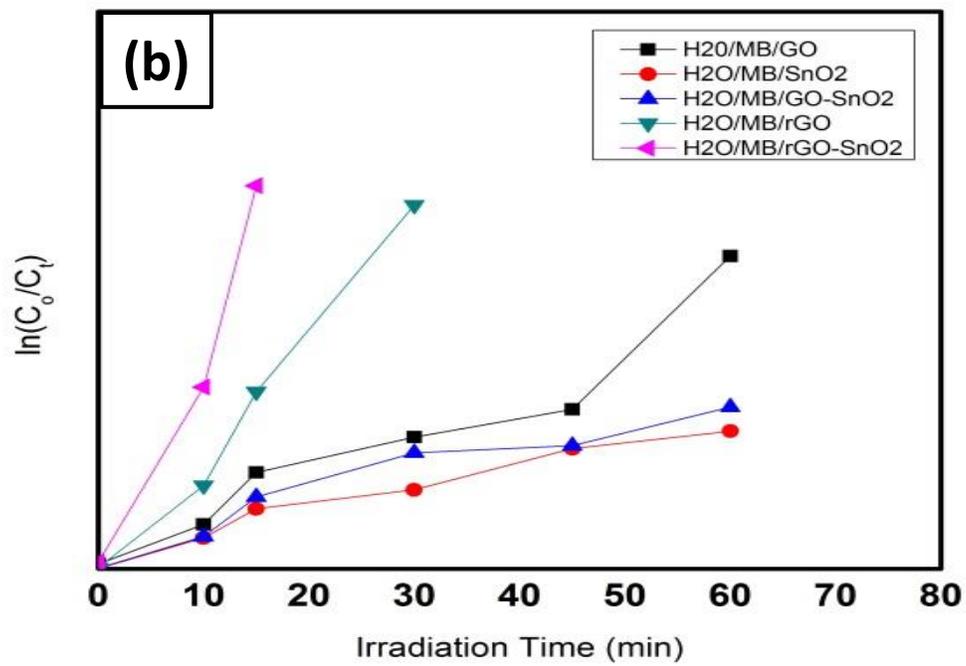

Fig. 6: (a) Absorbance versus Irradiation time curves for GO, $SnO_2$, GO-$SnO_2$, rGO and rGO-$SnO_2$ and (b) $\ln(C_0/C_t)$ versus Irradiation time curves illustrating MB photodegradation by GO, $SnO_2$, GO-$SnO_2$, rGO and rGO-$SnO_2$

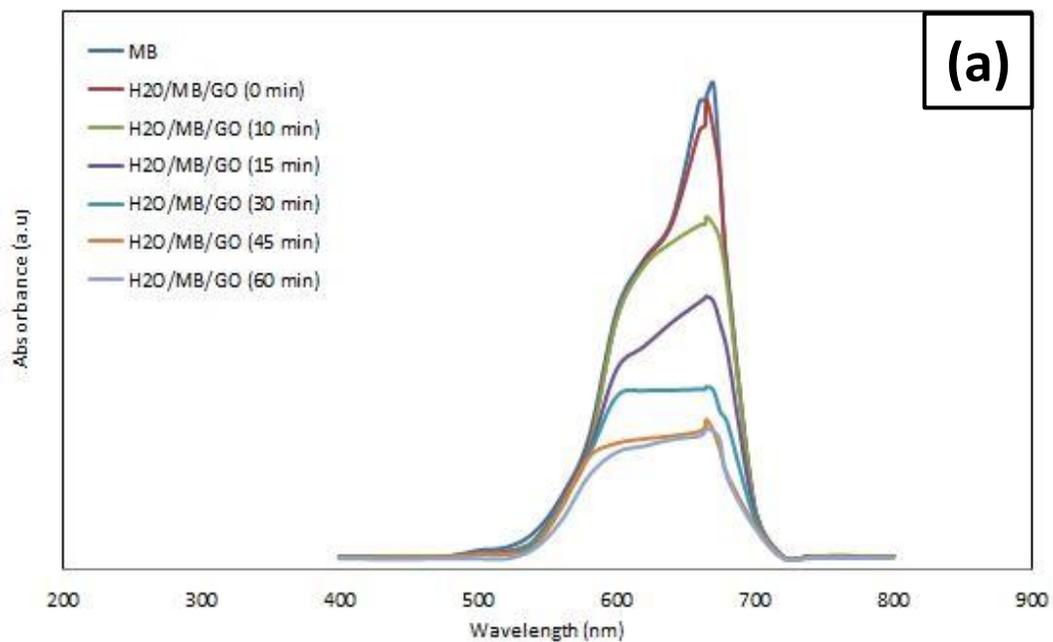

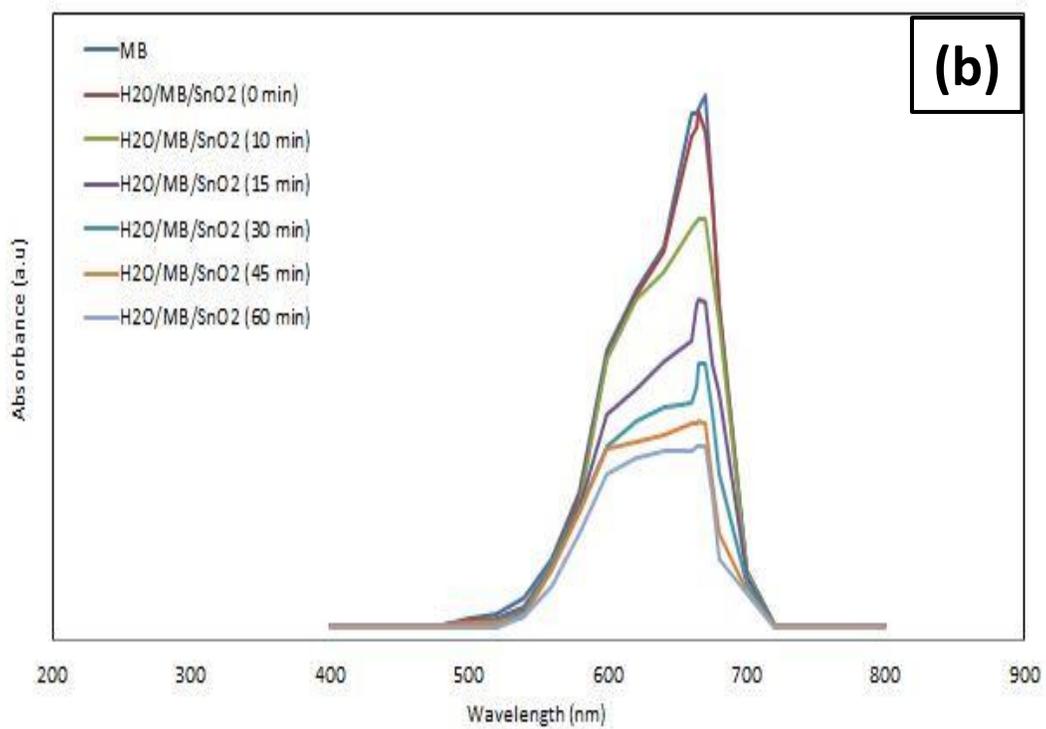

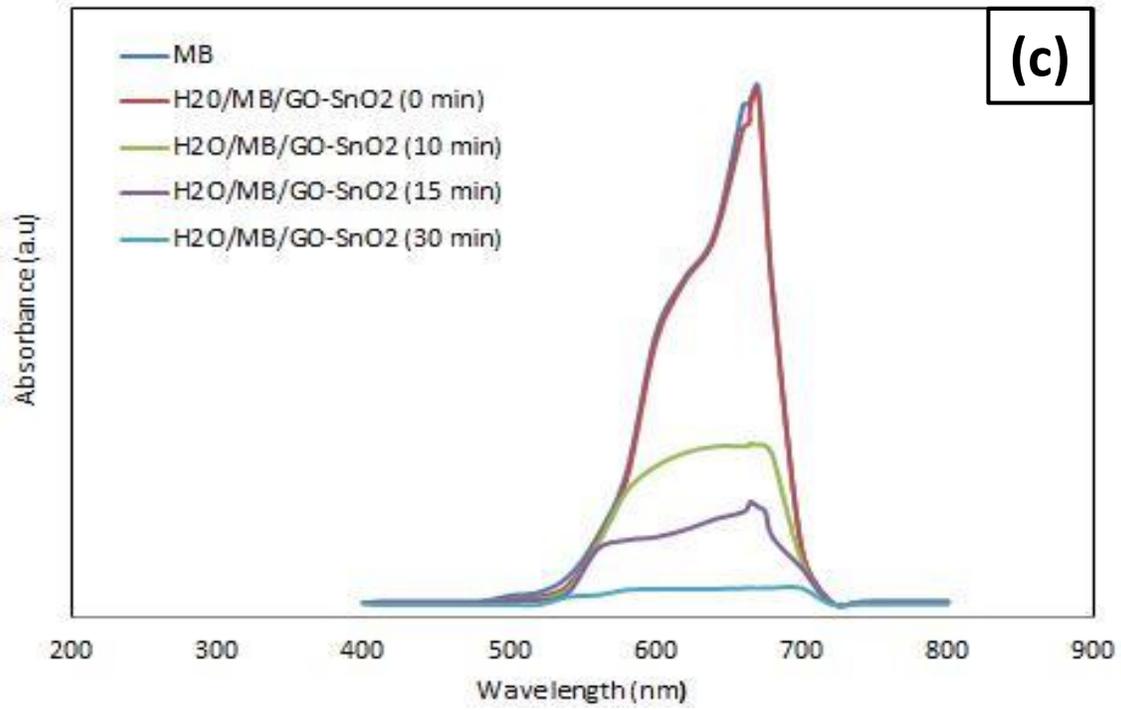
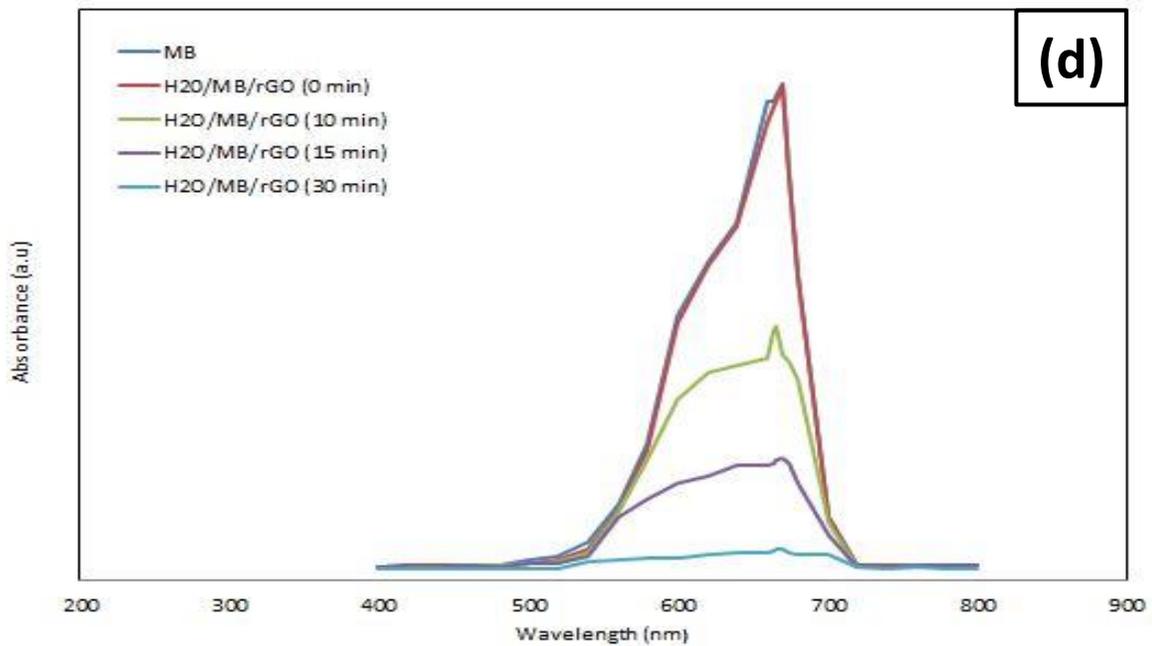

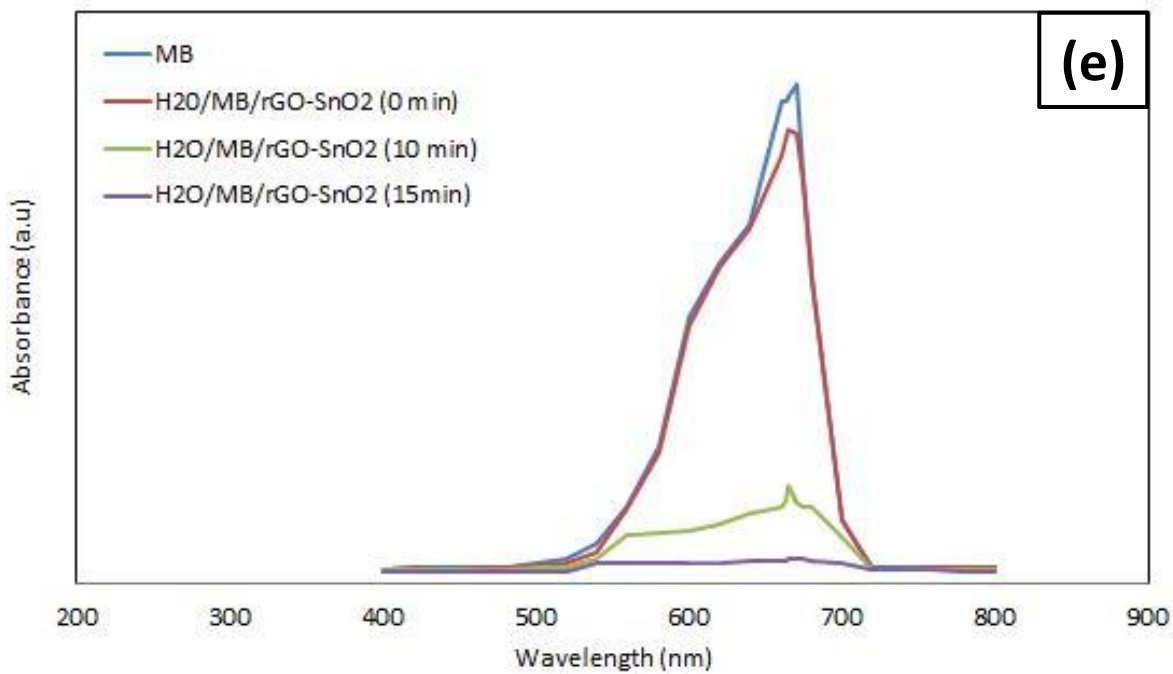

Fig. 7: Time-dependent absorption spectra of MB solution during UV light irradiation in the presence (a) GO, (b) $SnO_2$, (c) GO-$SnO_2$, (d) rGO and (e) rGO-$SnO_2$

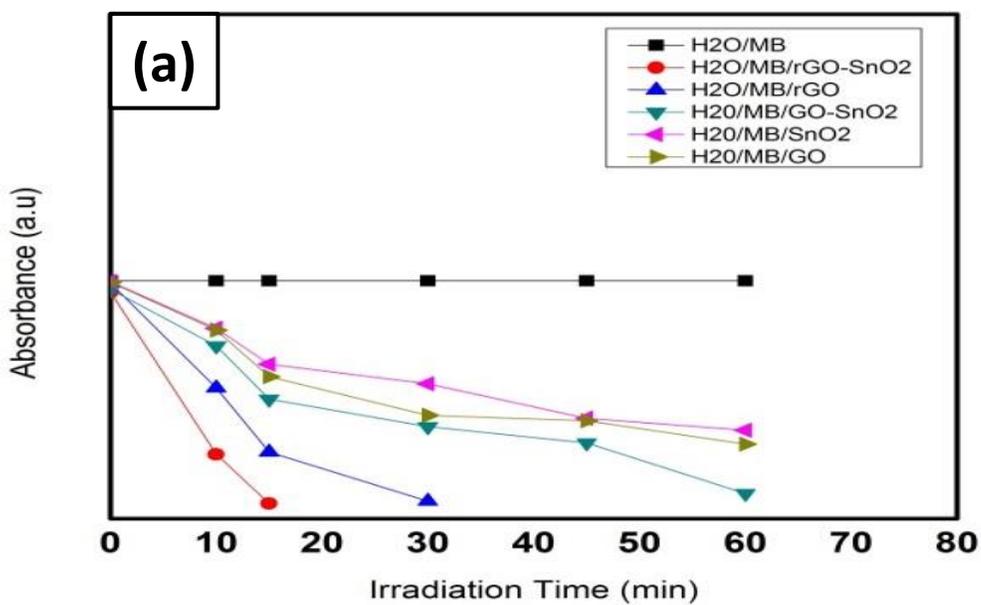

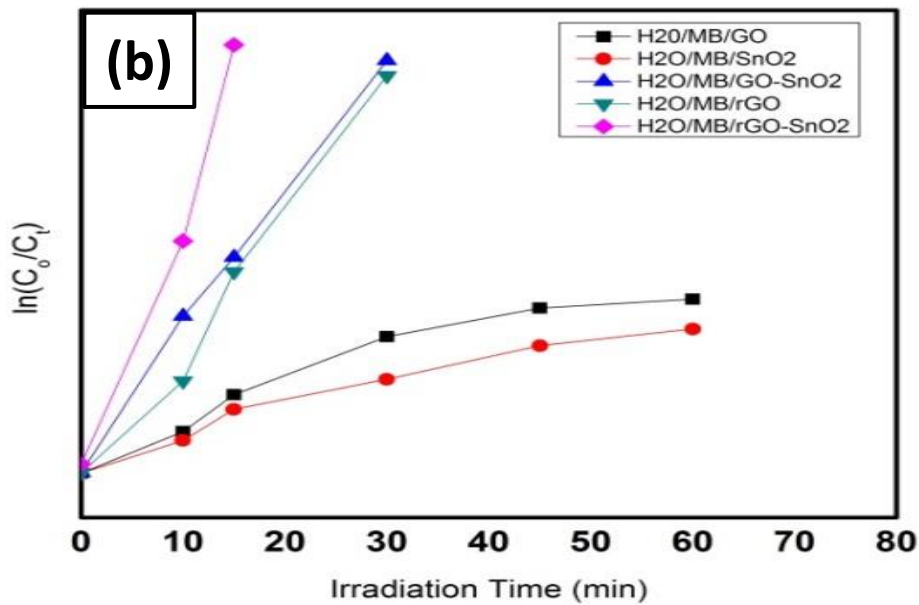

Fig. 8: (a) Absorbance versus Irradiation time curves for GO, $SnO_2$, GO-$SnO_2$, rGO and rGO-$SnO_2$ and (b) $\ln(C_0/C_t)$ versus UV light Irradiation time curves illustrating MB photo degradation by GO, $SnO_2$, GO-$SnO_2$, rGO and rGO-$SnO_2$

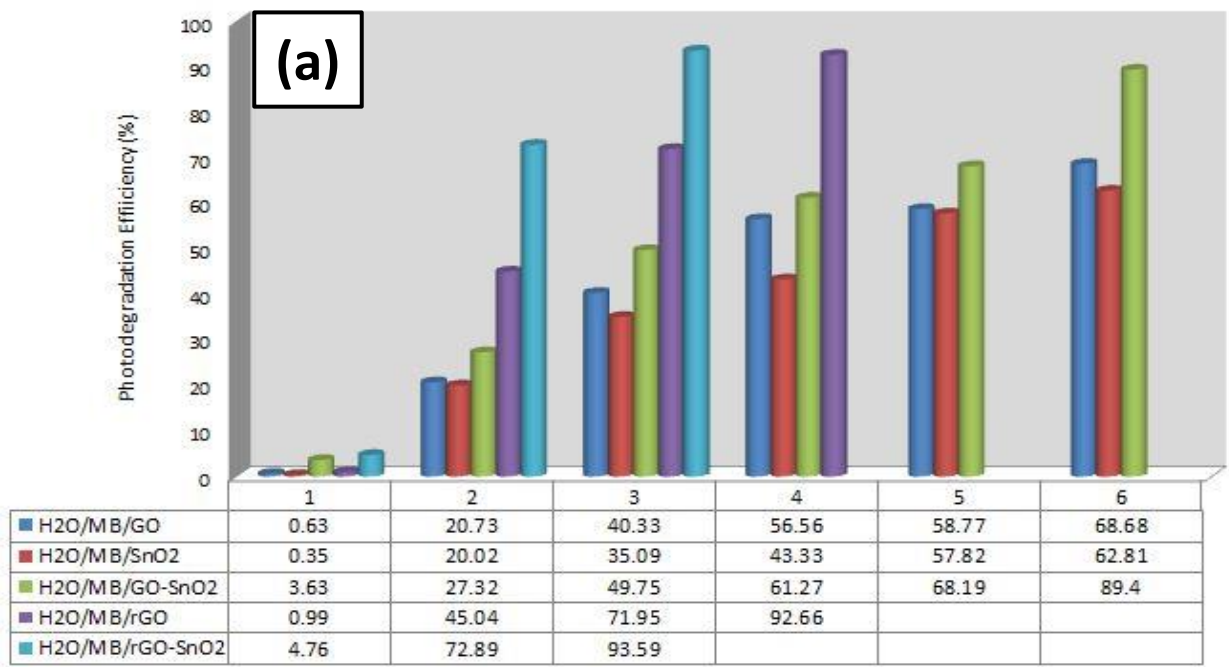

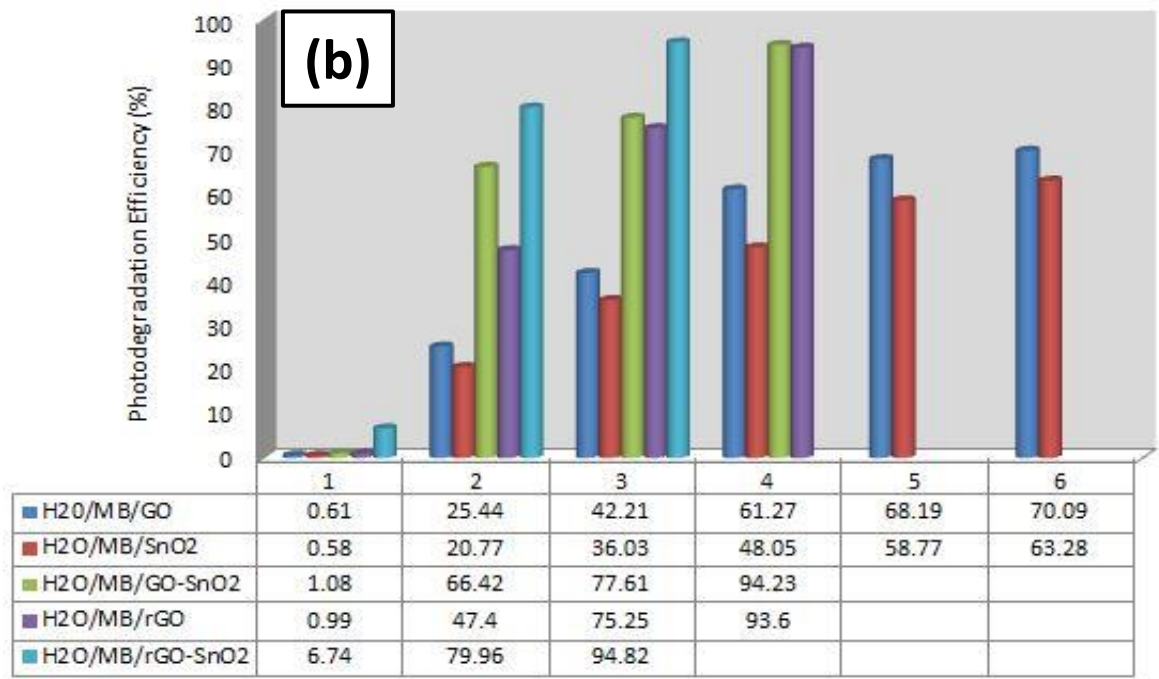

Fig. 9: Photodegradation Efficiency (%) under (a) Natural Sunlight and (b) UV light

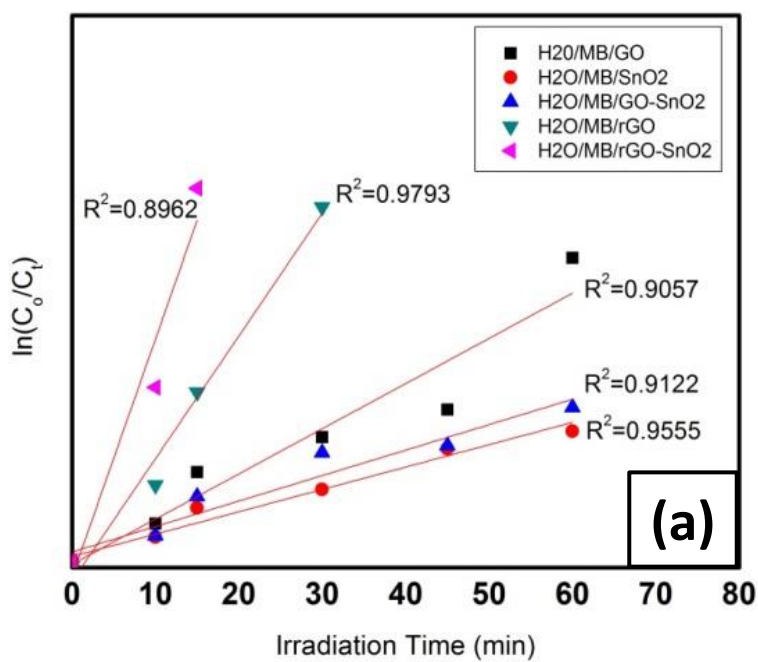

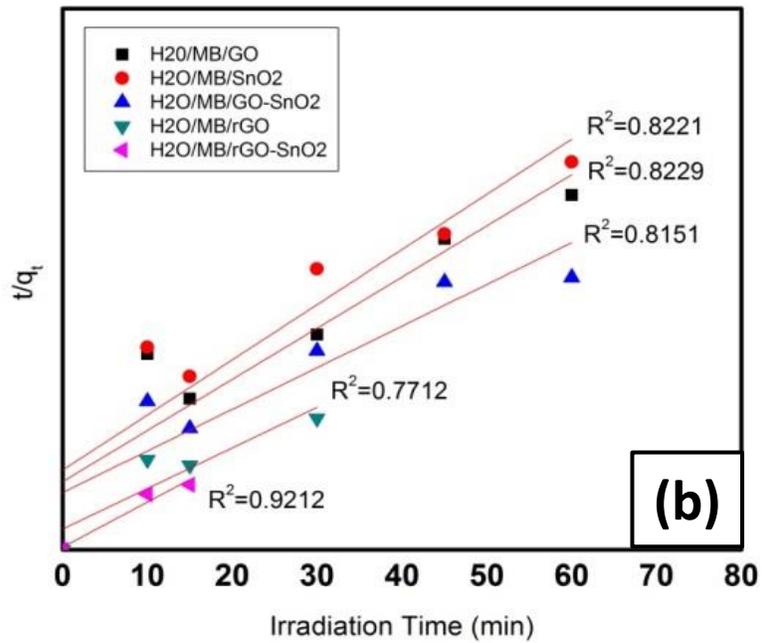

Fig. 10: The correlation coefficient ($R^2$) values for (a) pseudo-first order and (b) pseudo-second order under natural sunlight irradiation.

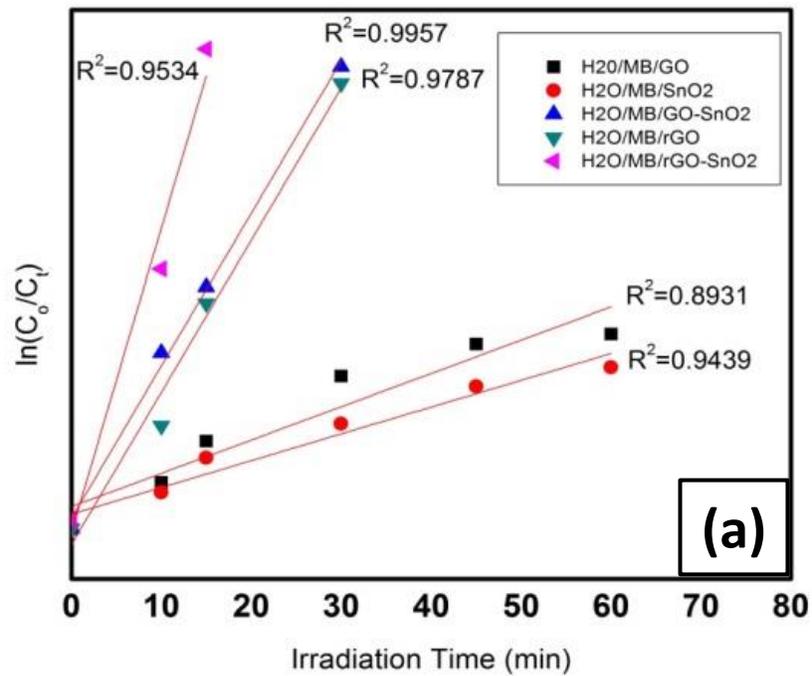

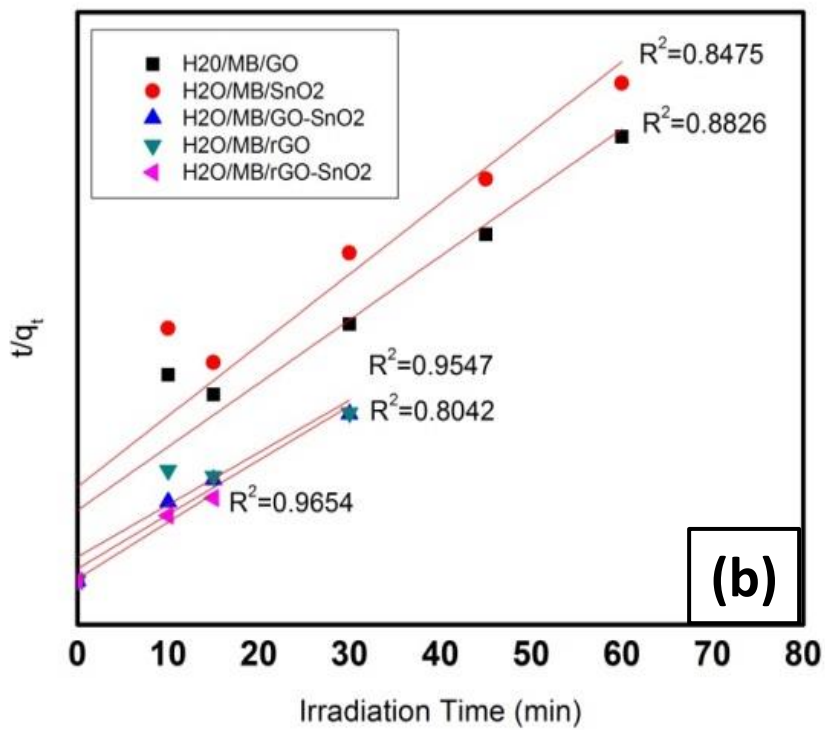

Fig. 11: The correlation coefficient ($R^2$) values for (a) pseudo-first order and (b) pseudo-second order under UV light irradiation.

# TABLES

### Table 1. Properties of Methylene Blue (MB).

| Properties | Cationic Azo Dye |
|---|---|
| Synonym name | Basic Blue 9 |
| Molecular formula | $C_{16}H_{18}ClN_3S$ |
| Molecular weight | 319.851 g/mol |
| Absorbance wavelength($\lambda_{max}$) | 664 nm |
| Molecular structure | 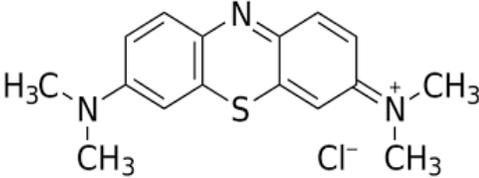 |

### Table 2. The photocatalytic degradation of MB using several catalysts.

| Authors/Year | Catalysts | Degradation efficiency (%) | Conditions | References |
|---|---|---|---|---|
| Soltani et al. 2014 | $BiFeO_3$ | 100% MB | Time: 80 min; Catalyst loading: 0.5 g/L$^{-1}$; Irradiation: Natural Sunlight; pH 2.5 | [5] |
| Dariani et al. 2016 | $TiO_2$ | 100% MB | Time: 2 hr; Catalyst loading: 0.5 g/L$^{-1}$; Irradiation: UV light; pH 2.5 | [6] |

Table 3. Degradation efficiency and pseudo-first order rate constant for photocatalytic degradation of MB by GO, SnO$_2$, GO-SnO$_2$, rGO and rGO-SnO$_2$ under natural sunlight irradiation:

| Samples | Concentration of MB (mM) | Degradation Efficiency (%) | $R^2$ | Rate constant (min$^{-1}$) |
|---|---|---|---|---|
| H$_2$O/MB/GO | 0.05 | 68.68% | 0.9057 | 0.0374 |
| H$_2$O/MB/SnO$_2$ | 0.05 | 62.81% | 0.9555 | 0.0165 |
| H$_2$O/MB/GO-SnO$_2$ | 0.05 | 89.4% | 0.9122 | 0.0193 |
| H$_2$O/MB/rGO | 0.05 | 92.66% | 0.9793 | 0.0870 |
| H$_2$O/MB/rGO-SnO$_2$ | 0.05 | 93.59% | 0.8962 | 0.1833 |

Table 4. Degradation efficiency and pseudo-second order rate constant for photocatalytic degradation of MB by GO, SnO$_2$, GO-SnO$_2$, rGO and rGO-SnO$_2$ under natural sunlight irradiation:

| Samples | Concentration of MB (mM) | Degradation Efficiency (%) | $R^2$ | Rate constant (mM$^{-1}$min$^{-1}$) |
|---|---|---|---|---|
| H$_2$O/MB/GO | 0.05 | 68.68% | 0.8229 | 0.3641 |
| H$_2$O/MB/SnO$_2$ | 0.05 | 62.81% | 0.8221 | 0.2816 |
| H$_2$O/MB/GO-SnO$_2$ | 0.05 | 89.4% | 0.8151 | 1.4059 |
| H$_2$O/MB/rGO | 0.05 | 92.66% | 0.7712 | 4.2079 |
| H$_2$O/MB/rGO-SnO$_2$ | 0.05 | 93.59% | 0.9212 | 9.75 |

**Table 5.** Degradation efficiency and pseudo-first order rate constant for photocatalytic degradation of MB by GO, $SnO_2$, GO-$SnO_2$, rGO and rGO-$SnO_2$ under UV light irradiation:

| Samples | Concentration of MB (mM) | Degradation Efficiency (%) | $R^2$ | Rate constant ($min^{-1}$) |
|---|---|---|---|---|
| $H_2O$/MB/GO | 0.05 | 70.01% | 0.8931 | 0.0201 |
| $H_2O$/MB/$SnO_2$ | 0.05 | 63.28% | 0.9439 | 0.0167 |
| $H_2O$/MB/GO-$SnO_2$ | 0.05 | 94.23% | 0.9957 | 0.0951 |
| $H_2O$/MB/rGO | 0.05 | 93.6% | 0.9787 | 0.0916 |
| $H_2O$/MB/rGO-$SnO_2$ | 0.05 | 94.82% | 0.9534 | 0.1974 |

**Table 6.** Degradation efficiency and pseudo-second order rate constant for photocatalytic degradation of MB by GO, $SnO_2$, GO-$SnO_2$, rGO and rGO-$SnO_2$ under UV light irradiation:

| Samples | Concentration of MB (mM) | Degradation Efficiency (%) | $R^2$ | Rate constant ($mM^{-1}min^{-1}$) |
|---|---|---|---|---|
| $H_2O$/MB/GO | 0.05 | 70.01% | 0.8826 | 0.3904 |
| $H_2O$/MB/$SnO_2$ | 0.05 | 63.28% | 0.8475 | 0.2872 |
| $H_2O$/MB/GO-$SnO_2$ | 0.05 | 94.23% | 0.8042 | 5.4437 |
| $H_2O$/MB/rGO | 0.05 | 93.6% | 0.9547 | 4.8750 |
| $H_2O$/MB/rGO-$SnO_2$ | 0.05 | 94.82% | 0.9654 | 12.2033 |

# GRAPHICAL ABSTRACT

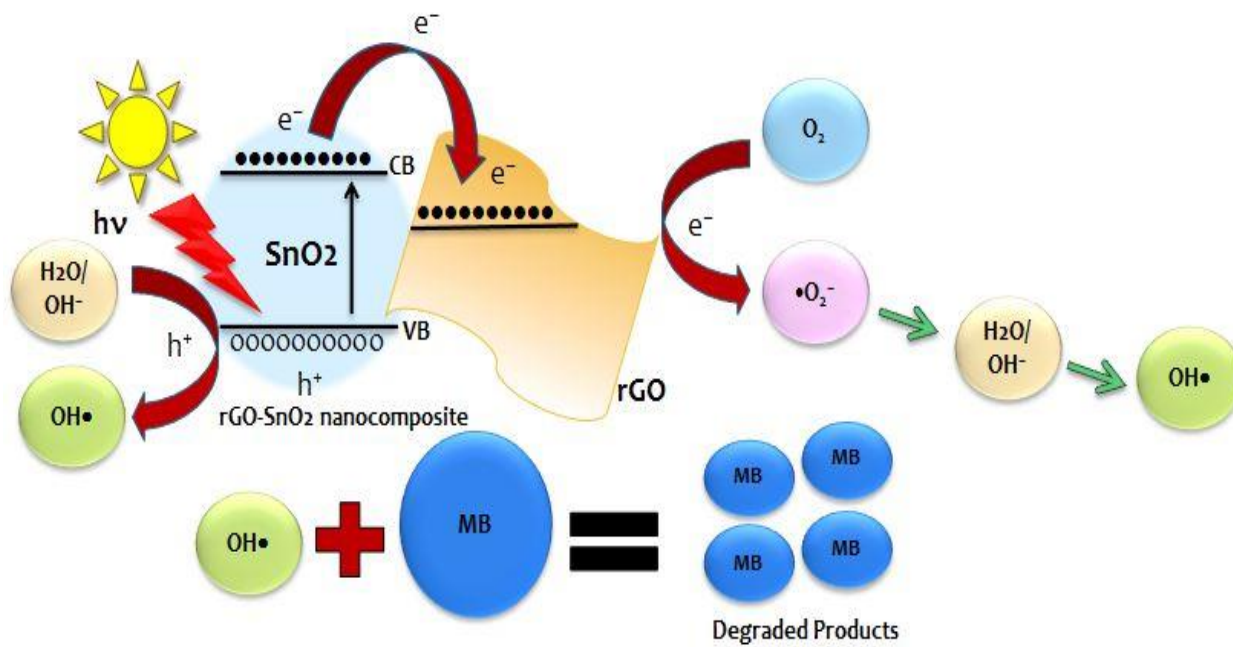

Fig. (A): Photodegradation of MB using rGO-SnO$_2$ nanocomposite under sunlight irradiation

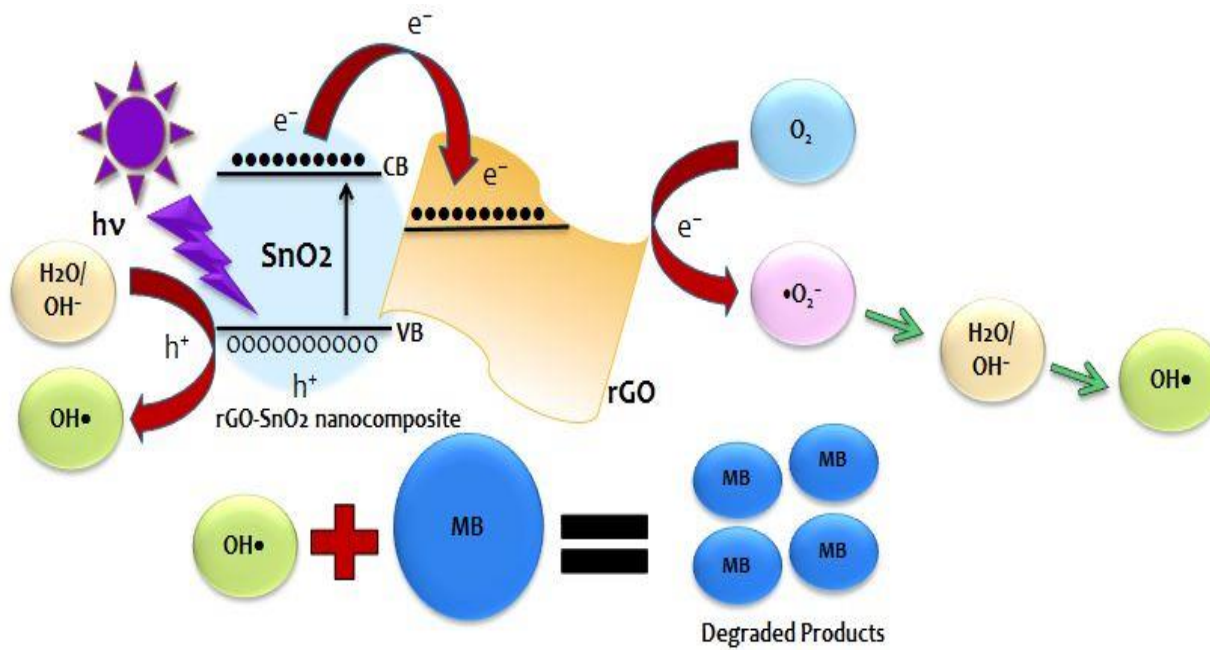

Fig. (B): Photodegradation of MB using rGO-SnO$_2$ nanocomposite under UV light irradiation